\documentstyle[12pt]{article} 
\def\vec#1{\mbox{\protect\boldmath $ #1 $}}

\begin{document}                                                              
\begin{flushright}
\begin{minipage}[t]{12em}
UAB-FT-459/98\\
IFT-P.087/98
\end{minipage}  
\end{flushright}
\vskip 1.5cm                                                      

\begin{center}                                                                
{ \large\bf BELL--INEQUALITIES FOR $K^0 \bar{K^0}$ PAIRS FROM 
$\Phi$--RESONANCE DECAYS }
\vskip 1.5cm  

{ B. Ancochea, A. Bramon 
} \\                 
Grup de F{\'\i}sica Te\`orica, Universitat Aut\`onoma de Barcelona, \\ 
08193 Bellaterra, Spain \\
\vskip 0.5cm
M. Nowakowski \\
Instituto de F{\'\i}sica Te\'{o}rica,
Universidade Estadual Paulista, \\
Rua Pamplona, 145 \\
01405 -900 S\~ao Paulo, Brazil \\
\end{center}

\vskip 3.5cm                             
\begin{abstract}
We analyze the premises of recent propositions to test local realism via
Bell--inequalities using neutral kaons from $\Phi$--resonance decays as
entangled EPR--pairs.  We pay special attention to the derivation of
Bell--inequalities, or related expressions, for unstable and oscillating 
kaon `quasi--spin' states and to the
possibility of the  actual identification of these states through their 
associated decay modes.  We discuss an indirect method to extract probabilities to
find  these states by combining experimental information with theoretical input. 
However, we still find inconsistencies in previous derivations of
Bell--inequalities.   
We show that
the identification of the quasi--spin states via their associated decay mode does
not allow the free choice to perform different tests on them,
a property which is crucial to establish the validity of any Bell--inequality 
in the context of local realism.
In view of this we propose a different kind of Bell--inequality in which 
the free choice or adjustability of the experimental set--up is guaranteed.
We also show that the proposed inequalities are violated by quantum mechanics. 
\end{abstract}                                                                

\newpage

\section{Introduction}

The quantum entanglement shown by the separate parts of a 
non--factorizable composite system is an extremely peculiar feature of
quantum mechanics and has recently become a powerful resource of new developments
like quantum teleportation and communication 
\cite{steane}. On the other hand, ever since the paper by 
Einstein, Podolsky and Rosen \cite{epr} quantum entanglement has been also a
continous source of food for thoughts and speculations on the ``spooky
action-at-a-distance'', better characterized as non-locality in the correlations of
an EPR--pair
\cite{peres}. A useful tool to probe into this non-locality has been given to 
us by Bell in form of his Bell--inequalities \cite{bell}. Known are also
related versions which a local realistic theory should satisfy, namely, 
the Clauser--Horne \cite{clh} and Wigner \cite{wigner} inequalities to which very
often we will simply and more generically refer as  Bell--inequalities. For a
general review on this subject we refer the reader to
\cite{review1}.

Bell--inequalities have been subjected to experimental tests with the general
outcome that they are violated \cite{bellexp}, i.e., local realistic theories
should be discarded and nature is indeed non--local. However, loopholes in the
tests have been pointed out \cite{santos}. It is then understandable that there
is a continous interest to test Bell--inequalities in different experiments
and, more importantly, in different branches of physics. One such place
which offers the opportunity to do exactly this is the $\Phi$--resonance, 
a $C=-1$ neutral vector meson decaying into $K^0 \bar{K^0}$ pairs. An $e^+e^-$
machine which is expected to produce a large amount of EPR-entangled 
$K^0\bar{K^0}$--pairs through the reaction
$e^+e^- \to \Phi \to K^0 \bar{K^0}$ will be soon operating in Frascati
\cite{handbook}. Due to the $C=-1$ nature of the $\Phi$--meson the
EPR entanglement of the neutral kaon pair can be explicitly written as
\begin{equation} \label{1}
|\Phi \rangle = {1 \over \sqrt{2}}\left[|K^0\rangle \otimes|\bar{K^0}\rangle
-|\bar{K^0}\rangle \otimes|K^0 \rangle \right]
\end{equation}
directly at the $\Phi$--decay point into the $K^0 \bar{K^0}$ initial state. 
The neutral kaons then fly apart allowing the definition by collimation of a
left  and a right hand  beam. Along these two beams kaon propagation takes 
places, including both $K^0 -\bar{K^0}$ oscillations and  $K_{L,S}$ weak
decays.
 
It would certainly add to our knowledge if we could examine the nature of
non-locality using unstable, oscillating states like these neutral kaons. 
Clearly, because of the non--trivial time development of the states involved here,
this scenario is quite distinct from the usually  considered Bohm reformulation of
EPR, even if the singlet--spin initial state,
\begin{equation} \label{2}
|0,0 \rangle = {1 \over \sqrt{2}}\left[|+ \rangle \otimes|- \rangle
-|- \rangle \otimes|+ \rangle \right],
\end{equation}
is formally identical to the initial $K^0 \bar{K^0}$ state (\ref{1}) and,  
moreover, in both cases one deals with an antisymmetric system consisting of 
two two-dimensional components. However, as will be evident below, the
derivation of the Bell--like inequalities for unstable, oscillating states requires
more  care and a too close analogy to the spin case can be misleading.

Early attempts to check local realistic theories in kaon $\Phi$--decays used
Bell--inequalities involving probabilities of $K^0$ and/or $\bar{K}^0$ detection 
defined at different times \cite{datta1}, \cite{datta2}, \cite{ghi}. 
The identification of $K^0$ {\it versus} $\bar{K}^0$ is not problematic and can
be performed exploiting their distinct strong--interactions on
nucleons. Moreover, the use of different detection times allows to fulfill 
a crucial prerequisite needed to derive Bell--inequalities from local realism, 
namely, that different measurements coresponding to alternative experimental 
set--ups could be alternatively performed over the measured system. 
Unfortunately, it was found that this kind of firmly derived Bell--inequalities
cannot be violated by quantum mechanics due to the specific values of kaon
parameters like their masses and decays widths (see
\cite{datta2} for similar investigations in the B--meson system). 

Recently, there has been a renewed interest in this subject 
\cite{domenico}--\cite{bertl}. The idea in \cite{domenico}--\cite{bf}
has been to drop the ``different time'' Bell--inequalities  
in favour of the identification of what are called `quasi--spin' kaon states. 
These are essentially arbitrary superpositions of the two $K^0$ and $\bar{K}^0$ 
basis states, $|K_{\alpha}\rangle = \alpha_1|K^0\rangle
+\alpha_2|\bar{K^0}\rangle$ with $|\alpha_1|^2 + |\alpha_2|^2 = 1$, defined 
in close analogy to the spinors in the spin case. The results look also quite
encouraging in the sense that one could show that these Bell--inequalities for
kaons are violated by quantum mechanics. However, as we will show in this paper,
there are several drawbacks in using this quasi--spin analogy. First of all the
identification of such quasi--spin kaon states is problematic (except for 
$K^0$ {\it vs} $\bar{K}^0$, as just mentioned) and has not been addressed in
full satisfaction so far. Indeed, it is not possible to observe such states
directly. An indirect method, using a theoretical input and direct observable
experimental  information, seems to be the only way to extract the probabilities to
find  specific quasi--spin states. We believe that we have found such a method,
which is an interesting result in its own respect. However, this is still not
sufficient to derive Bell--inequalities for these states in the context of local
realistic theories.  The reason is that the indirect identification
method mentioned above is only possible by a prior identification
(this is the experimental information) of decay products
of the unstable kaons. As previously stated, the theoretical derivation of any
Bell--inequality requires, as starting assumption, that the experimentalist has
the {\it free choice} among several different tests (like adjusting the 
different directions of spin--analyzers in the case of spin or the
correspondingly different detection times of $K^0$ and/or $\bar{K}^0$ in 
\cite{datta1}, \cite{datta2}, \cite{ghi}), the
other inherent properties of the states being fixed by the assumptions of local
realism. This is crucial to derive Bell--inequalities and is often forgotten in the
formalism. This possibility of experimental {\it intervention} is not guaranteed
if the quasi--spin is determined by identifying the decays products. We have
no possibilities of choice or intervention in this
situation neither on the decay time nor on the decay channel.
Therefore several of the suggested Bell--inequalities for neutral kaon pairs 
fail in their derivation from local realism which they are supposed to test
versus quantum mechanics.

In view of this unsatisfactory situation we propose a new 
type of Bell--inequali\-ties for entangled kaons produced in 
$\Phi$--resonance decays where the free choice
is indeed guaranteed in terms of the possibility of installing different
`regenerator' slabs along the kaon flight paths. These thin slabs
--which are essentially  those used in neutral-kaon regeneration experiments-- are
characterized by  adjustable parameters (like their thickness and nucleonic
density)  thus mimicking the different orientations of the 
analyzer in the analogous spin-case. Using such an experimental set--up,
we can then show that quantum mechanics predicts a violation of 
these Bell--inequalities.

The paper is organized as follows. In section 2 we discuss the 
possible kaon--state observables in $\Phi$--resonance decays. 
We show what condition
has to be put on the kaonic quasi--spin states in order to 
be able to extract the probabilities
to identify such states. A formula which determines such probabilities
from observable quantities is given.
In section 3 we discuss several Bell--inequalities previously 
suggested by other authors for entangled kaon pairs. We argue that in 
the analyses already performed reporting violations of the
inequalities by quantum mechanics, 
the experimental verification of this violation would not necessarily exclude
local realistic theories as the proposed inequalities are not a 
strict consequence of this kind of theories.
In section 4 we derive our new Bell--inequalities which do follow from local
realism  and show that they are violated by quantum mechanics. Section 5
summarizes our results.

\setcounter{equation}{0}
\section{Quasi--spin observables in neutral--kaon decays}

We start our discussion by quoting some elementary definitions regarding
neutral kaons and the properties and time evolution 
of $K^0$--$\bar{K}^0$ pairs from $\Phi$--resonance decays. 
The CP $=\pm 1$ eigenstates $K_{1/2}$ are defined by
\begin{equation} 
\label{2bis}
|K_{1/2}\rangle ={1 \over \sqrt{2}}\left[|K^0 \rangle \pm |\bar{K^0}
\rangle \right]
\end{equation}
and the mass eigenstates $|K_{S/L} \rangle $ in terms of $K_{1/2}$ 
and the CP--violation parameter $\epsilon$ are
\begin{eqnarray} \label{3}
|K_S\rangle &=&{1 \over \sqrt{1+|\epsilon|^2}}\left[|K_1\rangle
+\epsilon |K_2 \rangle \right] \nonumber \\
|K_L\rangle &=&{1 \over \sqrt{1+|\epsilon|^2}}\left[|K_2\rangle
+\epsilon |K_1 \rangle \right] 
\end{eqnarray}
The proper-time ($\tau$) development of these non--oscillating mass eigenstates
is given by
\begin{eqnarray} 
\label{3bis}
|K_{S/L}(\tau)\rangle &=&e^{-i\lambda_{S/L}\tau}|K_{S/L}\rangle
\nonumber \\
\lambda_{S/L}&=&m_{S/L}-{i \over 2}\Gamma_{S/L},
\end{eqnarray}
with $m_{S/L}$ and $\Gamma_{S/L}$ being the mass and width of $K_S$ and
$K_L$, respectively. This reverts into the corresponding time evolution for the 
initial two--component kaonic system (\ref{1})
\begin{equation} \label{4}
|\Phi(0)\rangle \to  
|\Phi(\tau_1, \tau_2)\rangle = {N \over \sqrt{2}}
\left[|K_S(\tau_1)\rangle \otimes|K_L(\tau_2)\rangle
-|K_L(\tau_1)\rangle \otimes|K_S(\tau_2) \rangle \right],
\end{equation}
with $|N|= (1+|\epsilon |^2) / |1-\epsilon ^2| \simeq 1$.
The arguments $\tau_1$ and $\tau_2$ refer to the proper times
of the time evolution on the left and right hand sides, respectively. 
For simplicity we restrict ourselves to the CPT-conserving case as the  
arguments we put forward are independent of any CPT-violation. 

Using eq.(\ref{4}), one can immediately construct a double decay amplitude
for weak kaon decays of the form \cite{peccei}
\begin{eqnarray} \label{5}
{\cal A}(f_1, \tau_1; f_2, \tau_2)&=&{1 \over \sqrt{2}}
\biggl [\langle f_1 |{   T}|K_S(\tau_1)\rangle \langle f_2 
|{   T}|
K_L(\tau_2)\rangle \nonumber \\
&-& \,\,\, \langle f_1 |{   T}|K_L(\tau_1)\rangle \langle f_2 
|{   T}|
K_S(\tau_2)\rangle \biggr ],
\end{eqnarray}
where $T$ is the transition operator and $f_i$ 
denotes the various $K_S$ and $K_L$ decay modes. 
The normalization of this doubly time dependent decay amplitude is such that
\begin{equation} \label{6}
\int_0^{\infty}d\tau_1\int_0^{\infty}d\tau_2\sum_{f_1 f_2}
|{\cal A}(f_1,\tau_1;f_2,\tau_2)|^2=1,
\end{equation}
where the summation $\sum_{f_1 f_2}$ includes also the phase space integrals
$\int dph(f_1), \,\int dph(f_2)$ which we define by
\begin{equation} \label{7}
\Gamma (K_{S/L} \to f)= \int dph(f)|\langle f
|{   T}|K_{S/L} \rangle |^2.
\end{equation}
Note that ${\cal A}$ in (\ref{5}) has different dimensions for different
n-body final states. It is then certainly useful to construct the {\it joint 
decay rate} $\Gamma (f_1,\tau_1;f_2, \tau_2)$ given by \cite{peccei}
\begin{equation} \label{8}
\Gamma (f_1, \tau_1;f_2, \tau_2) \equiv
{d^2 {\cal P} \over d\tau_1 d\tau_2}(f_1,\tau_1;f_2,\tau_2)
\equiv \int dph(f_1) \int dph(f_2)|{\cal A}(f_1,\tau_1;f_2,\tau_2)|^2
\end{equation}
This is a doubly time--dependent decay rate into both the specific decay
mode $f_1$ on the left beam at the time $\tau_1$ and $f_2$ on the right one 
at the time $\tau_2$. The way this joint decay rate --or {\it joint decay 
probability density} (in two times)-- is constructed makes it independent of the
momenta of the decay products and, much more  important for our discussion,
$\Gamma (f_1, \tau_1;f_2, \tau_2)$ is a standard, fully measurable quantity
at a $\Phi$--factory. 

Had we asked, at least formally at this stage, for an
observable to find a given kaon quasi--spin state 
$|K_{\alpha}\rangle \equiv \alpha_1|K^0\rangle
+\alpha_2|\bar{K^0}\rangle$ on the left and $|K_{\beta}\rangle$ (defined
correspondingly) on the right, we would have to calculate the 
{\it joint probability}
\begin{eqnarray} \label{9}
P(K_{\alpha},\tau_1;K_{\beta},\tau_2)&\equiv&
\biggl | {1 \over \sqrt{2}}[\langle K_{\alpha}|K_S(\tau_1)\rangle
\langle K_{\beta}|K_L(\tau_2)\rangle \nonumber \\
& & \,\,\, - \langle K_{\alpha}|K_L(\tau_1)\rangle
\langle K_{\beta}|K_S(\tau_2)\rangle ]  \biggr |^2
\end{eqnarray}
More precisely, $P(K_{\alpha},\tau_1;K_{\beta},\tau_2)$ is the probability 
of finding both an {\it undecayed} $K_{\alpha}$ on the left at $\tau_1$ and 
an {\it undecayed} $K_{\beta}$
on  the right at $\tau_2$ in a hypothetical experiment being also able to
distinguish between $K_{\alpha}$ and its orthogonal state $\tilde{K}_{\alpha}$ and 
between $K_{\beta}$ and $\tilde{K}_{\beta}$. It is a well--defined probability which
can be computed exactly in the same way as the corresponding one in the
spin-singlet case. Indeed, this latter too is obtained by simply projecting
eq.(\ref{2}) on the basis states defined by the spin-analyzer orientation. 
There are however three important differences: {\it i)} our 
$P(K_{\alpha},\tau_1;K_{\beta},\tau_2)$'s are not constant but (doubly)
time--dependent; {\it ii)} due to the kaon unstability, the probability 
normalizations are also different and a unifying `renormalization prescription'  
will be proposed at the end of this section; and, more importantly, {\it iii)}
in the spin case the probabilities are  directly measurable whereas in our kaon
quasi--spin  case they are not (except for ${K^0} - \bar{K^0}$), the  directly
measurable quantities being  the joint decay rates (\ref{8}). 

The convenience of working with these just defined joint probabilities 
\newline $P(K_{\alpha},\tau_1;K_{\beta},\tau_2)$ has  already been
noticed by other authors \cite{domenico}, \cite{bf}.
Benatti and Floreanini \cite{bf}, for instance,  
based their recent analysis on what they call ``the
`double decay probabilities' $P(f_1,\tau_1;f_2, \tau_2)$ , i.e., the
probabilities that one kaon  decays into a final state $f_1$ at proper  time
$\tau_1$, while the other kaon decays into the final state $f_2$ 
at proper  time
$\tau_2$". In their formalism (see below, \cite{bf} and \cite{remark1}) 
one has 
\begin{equation} \label{10}   
P(f_1,\tau_1;f_2,\tau_2)
={   Tr}\left[\left({\cal O}_{f_1}
\otimes {\cal O}_{f_2}\right)\rho_{\Phi}(\tau_1,\tau_2)\right],
\end{equation}
where $\rho_{\Phi}(\tau_1,\tau_2)$ is the density operator corresponding 
to the two--kaon state in eq.(\ref{4}) and ${\cal O}_{f_1}$ and 
${\cal O}_{f_2}$ are projector matrices describing each single kaon 
decay into $f_1$ and $f_2$ normalized by $Tr {\cal O}_{f_1} = Tr {\cal
O}_{f_2} = 1$. The same authors correctly stress that  their
$P(f_1,\tau_1;f_2,\tau_2)$ are {\it not} decay  rates and, 
in spite of calling them `double
decay probabilities', one can easily convince oneself 
(see also our analysis below leading
to eq.(\ref{23})) that the 
$P(f_1,\tau_1;f_2,\tau_2)$'s in ref. \cite{bf} defined by eq.(\ref{10})
coincide with our  $P(K_{f_1},\tau_1;K_{f_2},\tau_2)$'s 
defined by eq.(\ref{9}) once the
kaon  quasi--spin states $K_{\alpha}$ and $K_{\beta}$ are associated 
to the  specific decay
modes $f_1$ and $f_2$, respectively. The essential problem --a problem 
which is too naively
addressed in ref.\cite{bf} and not satisfactorily solved 
\cite{24}-- is then that these
theoretically  well-defined probabilities
$P(K_{f_1},\tau_1;K_{f_2},\tau_2) = P(f_1,\tau_1;f_2,\tau_2)$ 
are not directly measurable,
as we have already discussed. A relation between the latter  
probabilities and the truly
measurable decay rates $\Gamma (f_1, \tau_1;f_2, \tau_2)$ 
defined in eq.(\ref{8}) is
therefore highly desirable. 
A first  attempt along this direction has been proposed and
briefly discussed by  Di Domenico working in a similar context 
\cite{domenico}. However,
some   improvements are required to definitely establish such a relation as we 
proceed to discuss in the following paragraphs.

Our first step is to define the orthogonal basis containing a specific kaon 
state  associated to the physical (i.e., really accurring) $f$-decay mode
\begin{equation} \label{17}
|K_f \rangle \equiv {1 \over \sqrt{|a_f|^2 +|b_f|^2}}\left
[a_f |K_1 \rangle + b_f |K_2 \rangle \right ]
\end{equation}
and its orthogonal counterpart
\begin{eqnarray} \label{18}
|\tilde{K}_f \rangle & \equiv &{1 \over \sqrt{|\tilde{a}_f|^2 
+|\tilde{b}_f|^2}}\left
[\tilde{a}_f |K_1 \rangle + \tilde{b}_f |K_2 \rangle \right ],  
\end{eqnarray}
with $\langle K_f | \tilde{K}_f \rangle = 0$.
We fix the coefficients $\tilde{a}_f$ and $\tilde{b}_f$ by demanding
\begin{equation} \label{19}
\langle f |{   T}|\tilde{K}_f\rangle =0
\end{equation}
The unique solution (up to a phase) reads
\begin{eqnarray} \label{20}
|\tilde{K}_f\rangle &=&{\tilde{b}_f \over |\tilde{b}_f|}
{1 \over \sqrt{1 + |\tilde{r}_f|^2}}\left[\tilde{r}_f
|K_1 \rangle + |K_2 \rangle \right]\nonumber \\
|K_f\rangle &=&{a_f \over |a_f|}
{1 \over \sqrt{1 + |\tilde{r}_f|^2}}\left[
|K_1 \rangle -\tilde{r}_f^* |K_2 \rangle \right]
\end{eqnarray}
with
\begin{equation} \label{21}
\tilde{r}_f={\tilde{a}_f \over \tilde{b}_f}=-{\langle f
|{   T}|K_2\rangle
\over \langle f |{   T}|K_1 \rangle}
\end{equation}
It is now easy to check the following identities
\begin{eqnarray} \label{22}
& &|K_f \rangle \langle K_f |=\rho_{K_f}={\cal O}_f \nonumber \\
& &|\tilde{K}_f \rangle \langle \tilde{K}_f |=\rho_{\tilde{K}_f}=
{\cal O}_{\tilde f} \nonumber \\
& &|K_f \rangle \langle K_f |+ |\tilde{K}_f \rangle \langle 
\tilde{K}_f |={\cal O}_f +{\cal O}_{\tilde f}=1
\end{eqnarray}
From these equations and eqs.(\ref{9}) and (\ref{10}) one immediately obtains 
\begin{equation} \label{23}
P(f_1,\tau_1;f_2,\tau_2) ={   Tr}\left[{\cal O}_{f_1}\otimes
{\cal O}_{f_2})\rho_{\Phi}(\tau_1, \tau_2)\right] =
P(K_{f_1},\tau_1;K_{f_2},\tau_2) ,
\end{equation}
thus justifying the previously announced identification of 
$P(f_1,\tau_1;f_2,\tau_2)$ from ref. \cite{bf} with our
$P(K_{f_1},\tau_1;K_{f_2},\tau_2)$ in eq.(\ref{9}).
Note also that the last equation in (\ref{22}) is the correct unitarity sum 
for undecayed kaon states. 
 
Our second  step consists in expanding the physically decaying $K_S$ and $K_L$ 
states in two of the orthogonal bases just introduced: 
$K_{f_i}$ and ${\tilde{K}_{f_i}}$, with $f_i = f_1$ and $f_2$. One then has
\begin{eqnarray} \label{24}
|K_{S/L} \rangle &=& {1 \over \sqrt{|a_{S1/L1}|^2 + |\tilde{a}_{S1/L1}|^2}}
\left[a_{S1/L1}|K_{f_1}\rangle +\tilde{a}_{S1/L1}|\tilde{K}_{f_1}\rangle 
\right] \nonumber\\
|K_{S/L} \rangle &=& {1 \over \sqrt{|a_{S2/L2}|^2 + |\tilde{a}_{S2/L2}|^2}}
\left[a_{S2/L2}|K_{f_2}\rangle +\tilde{a}_{S2/L2}|\tilde{K}_{f_2}\rangle 
\right] 
\end{eqnarray}
Using equation (\ref{19}), we can now rewrite the double decay amplitude 
(\ref{5}) as follows
\begin{eqnarray} \label{25}
{\cal A}(f_1,\tau_1;f_2,\tau_2)&=&{1 \over \sqrt{2}}\biggl [
a_{S1}a_{L2}e^{-i\lambda_S \tau_1}e^{-i\lambda_L \tau_2}\langle
f_1
|{   T}|K_{f_1}\rangle \langle f_2
|{   T}|K_{f_2}\rangle \nonumber \\
& & \ \ \ - a_{L1}a_{S2}e^{-i\lambda_L \tau_1}e^{-i\lambda_S \tau_2}\langle
f_1 |{   T}|K_{f_1}\rangle \langle f_2 |{   T}|K_{f_2}\rangle \biggr ] .
\end{eqnarray}
Then, using eqs.(\ref{8}), (\ref{9}) and (\ref{25}) one can easily conclude that
\begin{eqnarray} \label{26}
& & \Gamma (f_1, \tau_1;f_2, \tau_2) = \nonumber \\
& &{1 \over 2}\left |a_{S1}a_{L2}e^{-i \lambda_S \tau_1}e^{-i\lambda_L \tau_2}-
a_{L1}a_{S2}e^{-i \lambda_L \tau_1}e^{-i\lambda_S \tau_2}\right |^2 
\Gamma (K_{f_1}\to f_1)\Gamma (K_{f_2}\to f_2) = \nonumber \\
& &P(K_{f_1}, \tau_1;K_{f_2},\tau_2)
\Gamma (K_{f_1}\to f_1)\Gamma (K_{f_2}\to f_2) ,
\end{eqnarray}
where
\begin{eqnarray} \label{27}
\Gamma (K_{f_1} \to f_1) &=& \int dph(f_1) |\langle K_{f_1}
|{   T}|f_1 \rangle
|^2 \nonumber \\
&=& \int dph(f_1)\left |a^*_{S1}\langle K_S |{   T}|f_1\rangle 
+a^*_{L1}\langle K_L|{   T}|f_1\rangle \right |^2
\end{eqnarray}
and the smallness of the  mass difference $\Delta m=m_S - m_L$ makes possible the
use  of the same phase--space factor for the two terms in the integrand of the
latter expression.

As a result of the algebraic manipulations in the last two paragraphs,  
we can now take  the joint decay rate
$\Gamma (f_1,\tau_1;f_2,\tau_2)$ from experiment and, quite independently,
we can also calculate 
$\Gamma (K_{f_{1/2}}\to f_{1/2})$ via eq.(\ref{27}). As announced before, 
one thus obtains the {\it joint probability} 
\begin{equation}\label{28}
P(K_{f_1},\tau_1;K_{f_2},\tau_2)={\Gamma (f_1,\tau_1;f_2,\tau_2)
\over \Gamma (K_{f_1}\to f_1)\Gamma (K_{f_2}\to f_2)}
\end{equation}
This equation is the desired connection between the simple and 
easily interpretable {\it joint probability}  
$P(K_{f_1},\tau_1;K_{f_2},\tau_2)$ and the measurable {\it joint decay rate} 
$\Gamma (f_1,\tau_1;f_2,\tau_2)$. A formally identical equation can be  found in
the analysis on the same topic performed by Di Domenico 
\cite{domenico}, but our expressions (\ref{27}) for $\Gamma (K_{f_1} \to f_1)$ and
the corresponding ones in \cite{domenico} are, unfortunately, not   
the same. Notice also that our procedure to extract the probability
$P(K_{f_1},\tau_1;K_{f_2},\tau_2)$ strongly relies on a theoretical input
in form of the condition (\ref{19}), where $f$ refers exclusively to 
physical, realistic $K_{S/L}$ decay modes. In other words, the same procedure
would not work for an {\it arbitrary} superposition of $K^0$ and
$\bar{K}^0$, because a relation, as established in (\ref{28}), between 
(\ref{8}) and (\ref{9}) does not hold in general.

Strictly speaking, this means also that probabilities such as 
$P(\tilde{K}_{f_1},\tau_1; K_{f_2}, \tau_2)$ or 
$P(\tilde{K}_{f_1},\tau_1; \tilde{K}_{f_2}, \tau_2)$, involving one or two kaon
states $\tilde {K}_{f} $, cannot be extracted by the same method. However, we can
do that in a different way using a certain approximation. Let us first introduce
the notion of `any', i.e., the probability to detect any of the two basis states on
one of the two sides and a specific state on the other. We have
\begin{eqnarray} \label{29}
P(-,\tau_1;K_{f_2};\tau_2)&\equiv& P(K_{f_1},\tau_1;K_{f_2},\tau_2)
+ P(\tilde{K}_{f_1},\tau_1;K_{f_2},\tau_2) \nonumber \\
&=& P(K^0,\tau_1;K_{f_2},\tau_2)
+ P(\bar{K}^0,\tau_1;K_{f_2},\tau_2),
\end{eqnarray}
where the bar `$-$' denotes that we have summed over the two possible
orthogonal outcomes on the left hand side. Similar definitions hold of course for
the right hand side and for both sides, i.e., $P(-,\tau_1;-\tau_2)$. It should be
clear that (\ref{29}) does not depend on the choice of $f_1$ on the left hand beam 
and therefore we can replace $K_{f_1}$ by $K^0$, as done in the second
line of (\ref{29}). In the excellent approximation of the $\Delta Q=\Delta S$ 
rule, we have $\langle \pi^+ l^- {\bar{\nu}}|{   T}|K^0 \rangle =0$ and
$\langle \pi^- l^+ \nu |{   T}|\bar{K}^0 \rangle =0$ thus fulfilling in both
cases a condition like that in eq.(\ref{19}). Thanks to this, both
probabilities in the second line in (\ref{29}) can be extracted via (\ref{28}). 
This obviously allows the subsequent computation of 
$P(\tilde{K}_{f_1}, \tau_1;K_{f_2},\tau_2)$ through eq.(\ref{29}). 
In other words, the basis consisting of the two strangeness eigenstates
$K^0$ and $\bar{K}^0$ is exceptional not only in that these two states can be 
unambiguously detected using their distinct strong interactions in nucleonic
matter but also in that eq.(\ref{28}) can be used to measure the $K^0$-- 
or $\bar{K}^0$--detection probabilities through their associated semileptonic decay
modes $\pi^- l^+ \nu$ or $\pi^+ l^- {\bar{\nu}}$, respectively. For another basis,
such as  that consisting of $K_f$ and $\tilde{K}_f$ associated to the $f$--decay
mode,  probabilities involving $K_f$--detection can be similarly measured via 
eq.(\ref{28}) but those for $\tilde{K}_f$--detection require the use of 
eq.(\ref{29}). Finally, for bases consisting of `quasi--spin' states not linked to
an specific, realistic decay mode none of the probabilities seems to be measurable.

Let us also mention that in order to formulate  
Bell--inequalities for unstable two-component systems like kaons, 
$P(K_{f_1}, \tau_1;K_{f_2},\tau_2)$ is not, strictly speaking, the most suitable
quantity.  The reason is exactly the unstability of the components under
consideration  which superimposes an irrelevant time evolution (due to weak
decays) to the  relevant one (due to quasi--spin oscillations). 
The suitable observable  for unstable
and oscillating states is not $P(K_{f_1},\tau_1;K_{f_2},\tau_2)$, but rather 
\begin{equation} \label{41}
p(K_{f_1},\tau_1;K_{f_2},\tau_2) \equiv 
{P(K_{f_1},\tau_1;K_{f_2},\tau_2) / P(-,\tau_1;-,\tau_2)}, 
\end{equation}
where 
\begin{eqnarray}\label{41b}
P(-,\tau_1;-,\tau_2) &=& 
P(K^0,\tau_1;K^0,\tau_2) + P(K^0,\tau_1;\bar{K}^0,\tau_2) + \nonumber  \\
& & P(\bar{K}^0,\tau_1;K^0,\tau_2) + P(\bar{K}^0,\tau_1;\bar{K}^0,\tau_2) 
\end{eqnarray}
is an obvious generalization of eq.(\ref{29}). 
Equation (\ref{41}) means that we have
``renormalized'' the probabilities not to the total number
of decay events, but to the restricted set of decays happening at the times
$\tau_1$ and $\tau_2$ and covering the four
possible outcomes associated to any given pair of dimension--two orthogonal bases,
as exemplified in eq.(\ref{41b}). 
From  eq.(\ref{41b}) one can easily compute
\begin{equation}\label{41c}
P(-,\tau_1;-,\tau_2) \simeq e^{-{1 \over 2}(\Gamma_L + \Gamma_S)(\tau_1 + \tau_2)}
{\cosh} \left[{1 \over 2}(\Gamma_L - \Gamma_S)(\tau_1 - \tau_2) \right],
\end{equation}
where small terms of order $|\epsilon|^2$ and higher have been safely neglected. 
This allows to cancel the spurious time evolution induced by decays in the 
$P(K_{f_1},\tau_1;K_{f_2},\tau_2)$'s defined by eq.(\ref{9}) and the
new $p(K_{f_1},\tau_1;K_{f_2},\tau_2)$'s turn out to be simply normalized by
$$
p(-,\tau_1;-,\tau_2) = p(K_{f_1},\tau_1;-,\tau_2)
+ p(\tilde {K}_{f_1},\tau_1;-,\tau_2) = 1 
$$
in such a way that the similarities between these $p(K_{f_1},\tau_1;K_{f_2},\tau_2)$'s
and the corresponding ones in the conventional spin case cannot be increased any
further. This renormalization is not an essential point in most applications of
Bell--inequalities for unstable systems
\cite{remark3}, but exceptions, which  without insisting on this point lead to
contradictions, can be shown to exist. 

\setcounter{equation}{0}
\section{Bell--inequalities for $K^0$-$\bar{K}^0$ systems in $\Phi$--decays}

In the last section we derived a formula (eq.(\ref{28})) and 
a `renormalization prescription' which yields the probability
$p(K_{f_1},\tau_1;K_{f_2},\tau_2)$ 
provided we have experimental information on the direct measurable quantity 
$\Gamma (f_1,\tau_1;f_2,\tau_2)$ defined in eq.(\ref{8}) and we impose
on the kaon states the crucial condition (\ref{19}) for the physical $f_{1,2}$
decay modes. It should be  noted that quite a lot of a theoretical input is
required to arrive at the probabilities $p(K_{f_1},\tau_1;K_{f_2},\tau_2)$. But,
apart from that, it might appear that these probabilities --so close to those
appearing in the spin case-- could be sufficient to establish well--defined
Bell--inequalities for $\Phi$--resonance  decays into neutral kaons. This is,
unfortunately, not the case.  To understand this point, we best  compare the
Bell--inequalities for the usually considered singlet--spin  state with the ones
suggested in \cite{domenico} and \cite{bf} for entangled kaon pairs.

Let $A = \vec{a}, \vec{a\prime},...$ ($B = \vec{b}, \vec{b\prime},...$) be the set of
the various directions among which we can {\it choose} to measure the polarization of
the spin one--half subsystem coming from the initial spin--singlet state (\ref{2})
and propagating along the left (right) hand beam. Let $s_i$, with
$s_i=\pm$ and $i=a,a\prime,b,b\prime...$, be the  various possible outcomes of these
measurements in units of $\hbar /2$. 
Following Redhead \cite{redhead}, any (i.e., deterministic or non-deterministic) local
realistic theory can be shown to satisfy the following equation 
\begin{equation} \label{30}
p(s_a,s_b,\lambda)_{\vec{a},\vec{b}} = 
p(s_a |\lambda)_{\vec{a}} p(s_b |\lambda)_{\vec{b}}
\rho(\lambda) ,
\end{equation}
where $p(s_a,s_b,\lambda)_{\vec{a},\vec{b}}$ refers to the joint probability for
the singlet (\ref{2}) to be emitted in a given state fully characterized by the set of
hidden variables $\lambda$ and to produce the outcomes
$s_a$ and $s_b$ when measuring the spin one--half projections along $\vec{a}$ and
$\vec{b}$. Obviously one also has $p(s_a,s_b,\lambda)_{\vec{a},\vec{b}} = 
p(s_a;s_b | \lambda)_{\vec{a},\vec{b}} \rho (\lambda)$. Here and 
in the right hand side of eq.(\ref{30}), the notation $p(X|Y)$ is reserved
for {\it conditional}  probabilities and $\rho (\lambda)$ is the probability
distribution for the two-component  system being emitted in the state $\lambda$ with
the obvious normalization $\int d\lambda \rho (\lambda)=1$. Equation (\ref{30}) 
--often referred to as `factorizability' rather than `locality' condition, as
discussed in detail in \cite{redhead} and \cite{lupus}-- is also
equivalent to the locality condition used by Clauser and Horne in 
\cite{clh} to derive their general class of Bell inequalities.

The derivation of these Bell--inequalities 
proceeds by requiring that the observed probabilities
correspond to an average of the $\lambda$--dependent probabilities
via
\begin{eqnarray} \label{34}
p(s_a;s_b)_{\vec{a},\vec{b}}
&=&\int d\lambda \rho (\lambda ) p(s_a|\lambda )_{\vec{a}}
p(s_b|\lambda )_{\vec{b}} \nonumber \\
p (s_a)_{\vec{a}}&=&\int d\lambda \rho (\lambda ) p(s_a |\lambda )_{\vec{a}}
\nonumber \\
p (s_b)_{\vec{b}}&=&\int d\lambda \rho (\lambda ) p(s_b |\lambda )_{\vec{b}}.
\end{eqnarray}
A general Bell--type inequality follows then from the
simple mathematical theorem stating that
$$
x_1x_2-x_1x_4+x_2x_3+x_3x_4 \leq x_3 + x_2 ,
$$
provided that $0 \leq x_i \leq 1$ \cite{clh}. Translating $x_ix_j$ into product
of probabilities 
\newline
$p(s_a |\lambda )_{\vec{a}} p(s_b |\lambda )_{\vec{b}}$, using then the
factorizability condition (\ref{30}) and finally integrating
over $\lambda$ one gets
\begin{equation} \label{35}
p(s_a;s_b)_{\vec{a},\vec{b}}- p(s_a;s_d)_{\vec{a},\vec{d}}+
p(s_c;s_b)_{\vec{c},\vec{b}}+ p(s_c;s_d)_{\vec{c},\vec{d}} \leq
p(s_c)_{\vec{c}} + p(s_b)_{\vec{b}} \,\, ,
\end{equation}
which is the well--known Clauser--Horne version of Bell--inequalities. 
 
Alternative versions of Bell--type inequalities 
can also be obtained. Possibly the most simple and best known  
is due to Wigner \cite{wigner}:    
\begin{equation} \label{36}
p(s_a;s_b)_{\vec{a},\vec{b}} \leq p(s_a;s_c)_{\vec{a},\vec{c}}+
p(s_c;s_b)_{\vec{c},\vec{b}}\,\,,
\end{equation}
which follows from identifying two of the four orientations in (\ref{35}) and 
requiring the perfect anticorrelation, $p(s_a;s_a)_{\vec{a},\vec{a}} = 0$, for the 
singlet state which is not only the obvious quantum mechanical prediction but 
also a well--tested experimental fact. This requirement, however, restricts the
derivability of Wigner--inequalities \cite{wigner} only to deterministic 
theories; indeed, if perfect anticorrelation is imposed in expressions analogous
to the first one in (\ref{34}) the various conditional probabilities 
$p(s_a|\lambda)_{\vec{a}}$ and $ p(s_b|\lambda)_{\vec{b}}$ turn out to be either
zero or one and, therefore, any stochastic local realistic theory collapses into a 
deterministic one (see \cite{redhead} for details).

Let us note here that a more detailed notation for $p(s_a)_{\vec{a}}$ 
(and, similarly, for $p(s_b)_{\vec{b}}$) would be
$p(s_a,\cdot)_{\vec{a},\vec{b}}$ where 
``$\cdot $'' denotes, as the bar 
``$\vec{-}$'' in section 2, that the two possible outcomes $s_b$ on the right hand side
have been integrated out. We can then write $p(s_a,\cdot)_{\vec{a},\vec{b}} 
\equiv p(s_a,+)_{\vec{a},\vec{b}}+p(s_a,-)_{\vec{a},\vec{b}}$.        
This makes then contact with the notation of section 2 which we will also
continue to use. The locality condition establishes that
$p(s_a)_{\vec{a}} =  p(s_a,\cdot)_{\vec{a},\vec{b}} = 
p(s_a,\cdot)_{\vec{a},\vec{b\prime}}$
is independent from the distant orientation $\vec{b},\vec{b\prime}$....     

In {\it purely formal} analogy to (\ref{35}) and (\ref{36}) we can now derive 
Bell--inequalities involving our previously discussed kaon identification
probabilities $p(K_{f_1},\tau_1;K_{f_2},\tau_2)$ (\ref{41}). Each probability 
$p(s_a;s_b)_{\vec{a},\vec{b}}$ in expressions (\ref{35}) and (\ref{36}) can be 
substituted by a corresponding  
$p(k_{f_1};k_{f_2})_{K_{f_1},\tau_1;K_{f_2},\tau_2}$, where the two dichotomic
arguments $k_{f_i}$ are assumed to take the values $k_{f_i}=+$ or $-$ according to
the identification of the `quasi--spin' state as $K_{f_i}$ or its orthogonal state 
$\tilde{K}_{f_i}$. Reverting to the notation introduced in section 2, one thus 
has $p(K_{f_1},\tau_1;K_{f_2},\tau_2) \equiv p(+;+)_{K_{f_1},\tau_1;K_{f_2},\tau_2}$,
$p(K_{f_1},\tau_1;\tilde{K}_{f_2},\tau_2)
\equiv p(+;-)_{K_{f_1},\tau_1;K_{f_2},\tau_2}$,... Using the shortest notation 
(quite in line with that in \cite{domenico} and \cite{bf}), the
Clauser--Horne inequality  (\ref{35}) can be rewritten as  
\begin{eqnarray} \label{38}
& &p(K_{f_1},\tau_1;K_{f_2},\tau_2) -
p(K_{f_1},\tau_1;K_{f_4},\tau_2) +
p(K_{f_3},\tau_1;K_{f_2},\tau_2) +
p(K_{f_3},\tau_1;K_{f_4},\tau_2) \leq \nonumber \\
& &p(K_{f_3},\tau_1;-,\tau_2) + p(-,\tau_1;K_{f_2},\tau_2) \,\,.
\end{eqnarray}
and a series of equivalent expressions obtained by replacing one or several $K_{f_i}$ by
$\tilde{K}_{f_i}$ consistently everywhere. 
The Wigner version of Bell--inequalities corresponding to (\ref{36}) 
restricts now to the equal time case, $\tau_1=\tau_1\equiv \tau$, and can be
immediately written as
\begin{equation} \label{39}
p(K_{f_3},\tau;K_{f_2},\tau) \leq
p(K_{f_3},\tau;K_{f_1},\tau) +
p(K_{f_1},\tau;K_{f_2},\tau) \,\, .
\end{equation}
As discussed in \cite{bf}, this simple expression follows also from the most
general  one (\ref{38}) by making the replacement $K_{f_1} \to \tilde{K}_{f_1}$ 
after having identified $f_1=f_4$, $\tau_1=\tau_1\equiv \tau$ and imposing 
$p(K_{f_1},\tau;K_{f_4},\tau) = 0$. 
But the previous considerations concerning this last requirement of perfect
anti--correlation  reduce the derivability of (\ref{39}) only to deterministic local
realistic theories \cite{redhead}.

One could argue that the {\it purely formal} analogy between the 
singlet--spin and the kaonic $\Phi$--decay cases discussed in 
the previous paragraph is broken by the different role 
played by the time parameter(s). This is only partially true. In both cases
time plays a fundamental role because the real riddle of the 
``spooky action-at-a-distance'' in quantum mechanical entanglement is the apparent
possibility of causally connecting {\it space-like} separated events. To make
sure that kaon decay events on the left  are causally disconnected from the events
on the right   we have to impose the condition
$x_1 + x_2 /|t_2 - t_1| \geq 1$,  where $x_1$ and $x_2$ are the distances travelled
along the left and right  sides, respectively. Using the semi-classical relation
$x=\beta c t$, which makes full sense to use for kaons from $\Phi$--decays
\cite{we} where $\beta \simeq 0.2$, we get
\begin{equation} \label{42}
{1 - \beta \over 1 +\beta }\leq {t_1 \over t_2}\leq
{1+\beta \over 1-\beta }
\end{equation}
which is symmetric in $t_1/t_2$.
For $t_1$ and $t_2$ obeying (\ref{42}) and, in particular, 
for $t_1 = t_2$ there cannot be any 
classical communication between the two events.
Equal times are then the most convenient choice and the two Wigner--inequalities 
(\ref{36}), where equal times are tacitly assumed, and (\ref{39}), where  
equal times are explicitly stated, are in perfect analogy. The situation is
different for the  Clauser--Horne inequalities (\ref{35}) and (\ref{38}), where the
explicit time  dependence of the latter allows for the {\it a priori} interesting
possibilities  first explored in \cite{datta1}, \cite{datta2}, \cite{ghi} (see, 
however, our comments below).

Having noticed some {\it purely formal} analogies we now turn to analyze a 
profound difference between the two cases we are considering.  
Whereas in the singlet--spin case the directions of the spin--analyzers 
can be adjusted at {\it free will} by the experimentalist who can choose among 
$A = \vec{a},\vec{a\prime},...$ and $B = \vec{b},\vec{b\prime},...$,
the decay mode in the kaonic case is not an observable 
we have any freedom to choose or adjust.
It is important to insist that this freedom to choose among different
tests to be eventually performed on the physical system is crucial in the
context of local realistic theories (see, e.g., \cite{bigi},  
\cite{redhead},\cite{peres2}, and our discussion above).  In these
theories, the behaviour of the physical system is contained in the set of its
hidden variables
$\lambda$. Whatever one {\it chooses} (provided a choice exists)
to measure produces an outcome which was somehow `inherent' in these hidden
variables  `instructions' telling the state how to react under each possible choice.
If alternative experimental measurements on a single system are admissible,
the corresponding probabilities for these  alternative choices with their different
possible outcomes are assumed to exist and a Bell--like inequality can in principle
be established in terms of these probabilities. However,  this is not possible if
there is no free choice on the side of the experimentalist, as happens when dealing
with decay modes and decay times   of freely propagating unstable particles. 
In other words, a particular kaon decay mode or decay instant is `contained'
already in the `set of instructions' parametrized by
$\lambda$ and, in general,  there is no possibility for a real choice allowing to
establish  Bell--inequalities. 

We are now in the position to pursue our discussion on the analyses recently
performed  by several authors trying to establish Bell--inequalities for entangled 
neutral-kaon pairs. We have reconsidered most of the arguments put forward  by 
Benatti and Floreanini \cite{bf} and, formally speaking, we have reached their same
generic Bell--inequalities (\ref{35}) and (\ref{36}). These authors then
concentrate on the Wigner--inequalities (\ref{36}) written also at equal times
and specified to  kaon quasi--spin states associated to the $\pi^+ \pi^-$, $\pi^0
\pi^0$ and $\pi^- l^+ \nu$ (or $\pi^+ l^- {\bar{\nu}}$, in a second inequality)
decay modes. Since the difference between the charged and neutral two-pion decay
amplitudes  is proportional to the phenomenological parameter 
$\epsilon \prime$ (which is a measure for direct CP--violation),
one  obtains the Bell--inequality $|Re  (\epsilon \prime)| \leq 
3 |\epsilon \prime|^2$. This inequality can clearly be violated by small (but
not  vanishing) values of  $\epsilon \prime$, which are quite compatible with
present day experimental data. Formally, we fully agree with all these results
found in \cite{bf} (see also \cite{remarkeps}). In our opinion,
however, the inequalities  in  \cite{bf} do not follow strictly from local
realistic theories: the  required possibility of intervention by the
experimentalist allowing a  choice among different measurements is not there if
one simply detects  decay events, as discussed in the  previous paragraph.
The same remark  applies to the detailed paper by di Domenico \cite{domenico}, 
where  similar Wigner--inequalities (not necessarily at equal times  here) are
also derived. The three binary alternatives proposed in this case, consist in 
identifying $K^0$ {\it vs} $\bar{K^0}$ (assuming the $\Delta S = \Delta Q$ rule  
for semileptonic decays), $K_1$ {\it vs} $K_2$ (via two-pion decays in the 
limit $\epsilon \prime = 0$) and a third quasi--spin state $\tilde K_S$ 
{\it vs} its orthogonal counterpart. The latter identification is achieved 
through a clever trick based on regeneration phenomena which has inspired 
our present treatment of the subject (see next section), but the
required possibility of choice is not contemplated.

To further convince the reader 
that there is real trouble in the Bell--inequalities proposed in these 
two papers let us also quote a previous analysis by Bigi 
\cite{bigi} in which the necessity of active choice or intervention by the
experimenter is explicitly emphasized. 
Indeed, the inequality proposed in  \cite{bigi} involving also three binary
alternatives is essentially the same as in the previous two analyses. However, 
the possibility of identifying the `third' quasispin direction, a possibility 
attempted only latter by Di Domenico \cite{domenico} and improved in the present
paper, is simply not  contemplated. Because of this, rather pessimistic
conclusions were reached in \cite{bigi}. Similar comments  apply to the recent
analysis by Uchiyama \cite{uchiyama}. Again, the requirement  of intervention by
the experimenter (choosing two measurements among three  possible options) is
stressed, but a new problem appears: the need of discriminating between the two
mass  eigenstates $K_S$ {\it vs} $K_L$. From 
the theoretical point of view, it is not obvious how to compute the corresponding
$K_S$ and $K_L$ detection probabilities since these two 
states are not orthogonal, $\langle K_S|K_L \rangle \neq 0$, due
to CP violation \cite{kabir}, \cite{enzo}. Indeed, naively computing these
probabilities by the usual quantum mechanical projections over $K_S$ or $K_L$
states can lead to paradoxa \cite{ab}, \cite{enzo}, 
\cite{datta3} and to curious effects \cite{kha}. Experimentally, 
discriminating $K_S$ from $K_L$ seems also not feasible and the possibility of 
deciding that we have a pure $K_L$ beam by waiting long enough until 
the `short' component died out (\cite{domenico}, \cite{eberhard}) would 
not work either in our case. Indeed, comparably large times, imposed by the 
space--like separation condition (\ref{42}), should be used also on the other side 
beam thus producing an almost complete  depletion of coincident counts. 

We have repeatedly argued above that the experimental  
violation of a Bell--inequality would  not necessarily signal a breakdown of local
realistic theories unless  the possibility of active intervention in the
corresponding experimental set up  is guaranteed. 
However, as explicitly indicated by the arguments of eq.(\ref{38}), 
each side of our neutral kaon EPR--configuration is characterized by:  {\it i)} a
kaon quasi--spin state $K_{f}$ (or its associated decay product $f$) and {\it ii)} a
time variable $t$ (or proper--time $\tau$).  Therefore the observables
entering these inequalities can be varied in another way. Indeed, in \cite{datta1},
\cite{datta2} and \cite{ghi} different {\it times} instead of different states were
used. Note that now the freedom of choice,  quite independent from the hidden
variables themselves, is indeed given in terms of the possibility of having
different $K^0$ {\it vs} $\bar{K}^0$ detection times. The Clauser--Horne
inequalities  following from the locality condition can be derived 
in the same way we reached at (\ref{35}). For instance and with an obvious
and simplified notation, they read 
\cite{datta1}
\begin{eqnarray} \label{43}
& &p(K^0,\tau_1;\bar{K}^0,\tau_2) - p(K^0,\tau_1;\bar{K}^0,\tau_4) + 
p(K^0,\tau_2;\bar{K}^0,\tau_3) + p(K^0,\tau_2;\bar{K}^0,\tau_4) \leq
\nonumber \\
& &p(-,\tau_1;\bar{K}^0,\tau_2) + p(K^0,\tau_1;-,\tau_2) \,\, 
\end{eqnarray}
and remain valid when replacing $K^0 \to\bar{K}^0$, or 
$\bar{K}^0 \to K^0$, or both $K^0 \leftrightarrow  \bar{K}^0$ (see also
\cite{datta2} for a generalization of (\ref{43})). It is worth pointing out
again that the states $K^0/\bar{K}^0$ are directly detectable through their
different strong interactions  on nucleon matter too. High--density detectors could
then be placed at conveniently choosen (time-of-flight) distances from the
production point.  Unfortunately it is then found in
\cite{datta1}  and \cite{datta2} that quantum mechanics does not violate these 
firmly established inequalities (\ref{43}) involving directly measurable
probabilities of finding $K^0-\bar{K}^0$ states. 

\setcounter{equation}{0}
\section{New Bell--inequalities for $K^0$-$\bar{K}^0$ systems in $\Phi$--decays}

In view of the discussions in sections 2 and 3, the situation of testing
quantum mechanics versus local realistic theories using 
$K^0-\bar{K}^0$ pairs from $\Phi$--resonance decays is quite unsatisfying. 
The inequality (\ref{43}) is a correct derivation of local realism, 
but as shown in \cite{datta1}--\cite{ghi}, quantum mechanics will not 
violate this inequality due to the specific values of the neutral kaon 
parameters. Hence, performing a discriminating test is not possible. Suggestions
like  those in \cite{domenico} and \cite{bf} have, in principle, two
drawbacks. One is the extraction from experiment of the  probabilities entering
the inequalities, the second one is the impossibility of having the required free
choice to perform different tests  aiming to identify the different quasi--spin
states
$K_f$.  Although the first point has already been clarified in section 2 and found 
that the  relevant probablities can be extracted in an indirect way, 
the second criticism still remains a defect of the suggested tests.
Other related suggestions use in their computations of the quantum mechanical
probabilities the projection method over $K_{S/L}$--states which, on account
of $\langle K_S|K_L \rangle \neq 0$, is not without ambiguities \cite{enzo}.
An asymmetric $\Phi$--factory is needed for other tests, as proposed in
\cite{eberhard}, but unfortunately such a factory will not be available in the
near future.

It is worth noting in this context that the $K^0-\bar{K}^0$ system
from $\Phi$--decays is one of the most interesting entangled
systems presently available to test quantum mechanics. 
We have here unstable and oscillating
states. In addition, this system is up to now the only system to display
CP--violation; indeed, the results in \cite{uchiyama} and \cite{bf} 
are seemingly related to the CP--violating parameters $\epsilon$ and
$\epsilon \prime$. It is therefore an interesting challenge to
search for a Bell--type inequality which on the one hand is a clear
consequence of local realism and on the other hand
could be violated by quantum mechanical predictions. Below we will
present such an inequality.

Instead of using different quasi--spin states $K_f$ in the probabilities, 
as in (\ref{38}) and (\ref{39}), or different times, as in (\ref{43}), 
we propose to exploit the possibilities that one has to modify 
by free choice the
propagation conditions along one  (or both) kaon flight path(s). This can be done
by introducing appropriate kaon `regenerators' or `absorbers', i.e., thin slabs of
nucleonic matter with adjustable  characteristics, which produce `quasispin
rotations' in the state of  the neutral kaons passing through. Such an `active
rotation' of the states has the same effects as changing the spin-analizer
orientation from, say, $\vec{a}$ to $\vec{a\prime}$ or counting $f$ rather 
than $f'$ decay modes. Over these modified states we then
need to detect kaon eigenstates only in a {\it single} quasispin direction, 
the most convenient one being  obviously that distinguishing $K^0$ from
$\bar{K}^0$. Indeed, these strangeness eigenstates can be identified both by their
distinct decay modes, as explained  in section 2, or by their different strong
interactions on nucleons in  a detector, as indicated above and explicitly
emphasized in \cite{ghi}.  As we can guarantee now a clearly free intervention of
the experimentalist  --who can adjust the parameters for different propagation
conditions translating into different `quasispin rotations'-- the resulting
Bell--inequalites reflect clearly the requirements and consequences  of local
realistic theories. However, the question whether quantum mechanics violates these
inequalities remains to be investigated.

In order to do this, we will restrict ourselves to the equal time situation
$\tau_1=\tau_2 \equiv \tau$ which ensures that the space-like separation of events is
automatically fullfilled. We can establish a complete analogy to the singlet--spin case
in form of the inequality (\ref{35})  --or, equivalently, (\ref{38})-- by writing
\begin{eqnarray} \label{44}
& &{p}(\kappa_1; \kappa_2)_{\nu_1, \nu_2} - 
{p}(\kappa_1; \kappa_4)_{\nu_1, \nu_4} +
{p}(\kappa_3; \kappa_2)_{\nu_3, \nu_2} +
{p}(\kappa_3; \kappa_4)_{\nu_3, \nu_4} \leq \nonumber \\\
& &{p}(\kappa_3; -)_{\nu_3} + {p}(- ;\kappa_2)_{\nu_2} \,\,\, ,
\end{eqnarray}
where ${p}$ is again a $\lambda$ averaged probability as before, $\kappa_i$ stands
for either $K^0$ or $\bar{K}^0$ detection and $\nu_i$ refers to the 
physical characteristics of the different absorbers that the experimentalist 
can introduce (or not, $\nu_i =0$) along the path(s). The Wigner version of 
Bell--inequalities can be obtained as before (see also \cite{we2}) 
\begin{equation} \label{45}
p(\kappa_1; \kappa_2)_{\nu_1 ,\nu_2 } \leq
p(\kappa_1; \kappa_3)_{\nu_1 ,\nu_3} + p(\kappa_3; \kappa_2)_{\nu_3 ,\nu_2} \,\, ,
 \end{equation}
which is simpler and less general than (\ref{44}), as previously discussed,
but it is also the most convenient for our elementary present purposes.

Before exploiting the inequality (\ref{45}), we have to examine briefly the  
regeneration of neutral kaons in homogeneous nucleonic media. 
We follow here \cite{domenico}, \cite{kabir} and \cite{ab}, where further details
can be found. The eigenstates of the mass matrix inside nucleonic matter are
\begin{eqnarray} \label{46}
|K_S'\rangle \simeq |K_S \rangle -\varrho |K_L \rangle
\nonumber \\
|K_L'\rangle \simeq |K_L \rangle +\varrho |K_S \rangle,
\end{eqnarray}
where we have neglected (small) corrections of order $\varrho^2$ and higher. 
This crucial regeneration parameter, $\varrho$, is defined as
\begin{equation} \label{47}
\varrho ={\pi \nu \over m_K}{f-\bar{f} \over \lambda_S-\lambda_L},
\end{equation}
where $m_K=(m_S + m_L)/2$, $f$($\bar{f}$) is the forward scattering
amplitude for $K^0$($\bar{K}^0$) on nucleons and $\nu$ is the nucleonic density 
(for numerical values and a detailed discussion, see \cite{domenico}).
This is probably the easiest parameter to adjust in an experimental 
set up, hence its explicit appearance in our notation for 
the inequalities (\ref{44}) and 
(\ref{45}). The time evolution inside matter for the eigenstates
$|K_{S/L}'\rangle$  follows the standard exponential and non--oscillating form
\begin{equation} \label{48}
|K_{S/L}'(\tau)\rangle
=e^{-i\lambda_{S/L}'\tau}|K_{S/L}'\rangle,
\end{equation}
where
\begin{eqnarray} \label{49}
\lambda_{S/L}'&\simeq& \lambda_{S/L}-\Delta \lambda +{\cal O}(\varrho^2)
\nonumber \\ 
\Delta \lambda&=& {\pi \nu \over m_K}(f+\bar{f})
\end{eqnarray}
This allows to compute the net effect of a thin absorber over the 
entering  $|K_{S/L}\rangle$ states in three steps: {\it i)} using eq.(\ref{46}) the
entering $|K_{S/L}\rangle$ states are projected into the $|K_{S/L}'\rangle$--basis 
which is the appropriate to account for inside matter propagation,
{\it ii)} the inside matter time--evolution of the latter is then taken into
account as dictated by eq.(\ref{48}) and, finally, {\it iii)} one reverts to the 
original $|K_{S/L}\rangle$--basis using again eq.(\ref{46}). One thus finds 
(see, for instance, \cite{domenico} and \cite{kabir})
\begin{eqnarray} \label{50}
|K_S\rangle \to e^{-i\lambda_S'\Delta \tau}
\left(|K_S \rangle +i \varrho (\lambda'_S - \lambda'_L )\Delta \tau |K_L \rangle \right)
\simeq |K_S \rangle + \eta (\varrho) |K_L \rangle  \nonumber \\
|K_L\rangle \to e^{-i\lambda_L'\Delta \tau}
\left(|K_L \rangle +i \varrho (\lambda'_S - \lambda'_L )\Delta \tau |K_S \rangle \right)
\simeq |K_L \rangle + \eta (\varrho) |K_S \rangle  ,
\end{eqnarray}
where $\Delta \tau$ is the time-of-flight inside matter (short enough to justify 
the use of first order approximations) and $\eta (\varrho) \equiv 
i\varrho (m_S -m_L)\Delta \tau + (1/2)\varrho (\Gamma_S -\Gamma_L)\Delta \tau$.
 
To calculate the probabilities appearing in eq.(\ref{45}) we need also the time
development of the initial entangled pair in (\ref{1}) or, more precisely, 
in (\ref{4}) referring to the $|K_{S/L}\rangle$ free--propagating states. 
Let us consider a  symmetric situation in which the kaons move in vacuum up 
to a proper time $\tau_1$ on both sides. At this time $\tau_1$, one kaon enters the
absorber we put on the left hand side (the parameters of this absorber will be
distinguished by a prime) and simultaneously the other kaon enters a right hand
side absorber (with parameters denoted by a double prime). If we follow now the
time evolution of the entangled kaon pair up to the total exit time, $\tau= \tau_1 +
\Delta \tau$, we get in our usual thin absorber approximation
\begin{eqnarray} \label{51}
& &|\Phi ( \tau, \varrho'; \tau,\varrho'')\rangle \simeq \\
& &{N(\tau) \over \sqrt{2}} 
\left[|K_L\rangle \otimes |K_S
\rangle -  |K_S \rangle \otimes |K_L\rangle
+\eta(\varrho',\varrho'')
\left(|K_L\rangle \otimes |K_L \rangle - 
|K_S \rangle \otimes |K_S\rangle
\right)\right] \simeq \nonumber \\
& &{N(\tau) \over \sqrt{2}}
\left[|\bar{K}^0\rangle \otimes |K^0 \rangle - 
|K^0 \rangle \otimes |\bar{K}^0\rangle
+\eta(\varrho',\varrho'')
\left(|K^0\rangle \otimes |\bar{K}^0 \rangle + 
 |\bar{K}^0\rangle \otimes |K^0\rangle
\right)\right],\nonumber
\end{eqnarray}
where, apart from a global phase, $|N(\tau)| \equiv  (1+|\epsilon |^2) 
e^{- {1 \over 2} (\Gamma_S + \Gamma_L) \tau} / |1- \epsilon^2|$ and 
\begin{equation} \label{52x}
\eta(\varrho',\varrho'')\equiv -i(\varrho''-\varrho')
(\lambda_L-\lambda_S)\Delta \tau .
\end{equation}
The cases with only one absorber on one of the two sides can be
recovered from (\ref{51}) by letting one of the $\varrho'$ or
$\varrho''$ go to zero.

Let us now concentrate on two specific versions of the inequality
(\ref{45}), namely, 
\begin{eqnarray} \label{52}
p(K^0 ;\bar{K}^0)_{0,\nu} &\le&
p(K^0 ;\bar{K}^0)_{0,2\nu}+ p(\bar{K}^0 ;\bar{K}^0)_{2\nu,\nu}
\nonumber \\ 
p(K^0 ;\bar{K}^0)_{0,\nu} &\le&
p(K^0 ;{K}^0)_{0,2\nu}+ p({K}^0 ;\bar{K}^0)_{2\nu,\nu}
\end{eqnarray}
where the two arguments refer to the particles detected and the 
subindices correspond to the absence of an absorber $(\nu_i =0)$, 
to its presence  $(\nu_i =\nu)$ and to the presence of a double 
density absorber $(\nu_i = 2 \nu)$.
Using (\ref{51}), the first inequality in (\ref{52}), leads to
\begin{equation} \label{53}
2\, \Re e \left[\eta (0,\varrho (\nu))\right] \le 0,
\end{equation}
whereas the second one gives
\begin{equation} \label{54}
0\le 4\, \Re e \left[\eta (0, \varrho (\nu))\right].
\end{equation}
Clearly we have achieved our objectives, at least at the most simple level. The
Bell--inequalities (\ref{45})  follow from deterministic local realism and one of
its two possible versions is predicted to be violated by quantum mechanics. Note
that this eventual violation should be there for any absorber and, less
importantly, also  independent of the small CP--violating parameters. Of course, it
remains to analyze how such a tiny violation of (\ref{45}) can be increased to a
finite, observable level and to check whether it is confirmed by the experiment or not.

\setcounter{equation}{0}
\section{Conclusions}

In this paper we have investigated possible tests of local realism through
Bell inequalities using $\Phi$--resonance decays into entangled neutral kaon
pairs. For previously suggested Bell--inequalities, one finds that either they are 
not violated by quantum mechanics (which renders any test impossible) or the
inequality itself could not be considered as a strict consequence of 
local realism. 

As far as the latter is concerned, we could clarify how to extract
the probabilities entering these inequalities from experiments performable with 
$\Phi$--resonance decays. It turned out that this is not possible for 
arbitrary quasi--spin kaon states, but only for specially defined
$K_f$ associated to physically occurring decay modes $f$. This in its own right is
an interesting observation which might have some consequences in considering
future test using $\Phi$--decays into two kaons. However, the impossibility of
submitting the kaon states to different identification tests and the necessity of
having to identify the kaonic states $K_f$ by its associate decays mode (on
which one has no possibility of intervention or choice) excludes these inequalities
to be considered a `true' Bell--inequality in the local realistic
sense. 

To improve this situation, we therefore suggest a new experimental configuration
associated to Bell--inequalities having all the virtues like (i) being strictly derived
from local realism and (ii) being violated by quantum mechanics regardless the
parameters of the system. 

\vskip 1cm

{\bf Acknowledgments}. This work has been partly 
supported by the Spanish Ministerio de Educaci\'on y Ciencia, by the EURODAPHNE 
EEC-TMR program CT98-0169 and by
Funda\c{c}\~ao de Amparo \`a Pesquisa do Estado de S\~ao Paulo (FAPESP)
and Programa de Apoio a N\'ucleos de Excel\^encia (PRONEX).


\begin{thebibliography}{99}
\bibitem{remark3}
For instance, when inserting quantum mechanical predictions in the
Bell--inequalities of the next section, 
$P(-,\tau_1;-\tau_2)$ becomes a common denominator and can be dropped.
\end{thebibliography}
\end{document}